\documentclass[%
 reprint,
 twocolumn,
 amsmath,amssymb,
 aps,
]{article}
\usepackage{amsmath}
\usepackage[marginal]{footmisc}
\usepackage{verbatim}
\usepackage{graphicx}
\usepackage{dcolumn}
\usepackage{bm}
\usepackage{multirow}
\usepackage{authblk}

\begin{document}

\title{Sensitivity of $CP$ Violation of $\Lambda$ decay in $J/\psi \to \Lambda \bar{\Lambda}$ at STCF\\}

\author[1]{Yue Xu\thanks{These authors contributed equally to the work.}}
\author[2,3]{Xiaorong Zhou$^*$}
\author[2,3]{Xiaodong Shi}
\author[1]{Yongxin Guo}
\author[1]{Kuiyong Liu}
\author[1]{Li Gong}
\author[1]{Xiaoshen Kang\thanks{Corresponding author.}}
\affil[1]{Department of Physics, Liaoning University, Shenyang 110036, People’s Republic of China}
\affil[2]{Department of Modern Physics, University of Science and Technology of China, Hefei 230026, People's Republic of China}
\affil[3]{State Key Laboratory of Particle Detection and Electronics, Hefei 230026, People’s Republic of China}


\date{}
\maketitle

\begin{abstract}
The process of $J/\psi \to \Lambda \bar{\Lambda}$ is studied using $1.0\times10^{12}$ $J/\psi$ Monte Carlo~(MC) events at $\sqrt{s}$=3.097~GeV with a fast simulation software at future Super Tau Charm Facility~(STCF). The statistical sensitivity for $CP$ violation is determined to be the order of $\mathcal{O}~(10^{-4})$ by measuring the asymmetric parameters of the $\Lambda$ decay. Furthermore, the decay of $J/\psi \to \Lambda \bar{\Lambda}$ also serves as a benchmark process to optimize the detector responses using the interface provided by the fast simulation software. 


\end{abstract}

\section{Introduction}

The electromagnetic force, weak nuclear force, and strong nuclear force are addressed with the Standard Model~(SM), which is established as a well-tested physics theory. Although SM is so successful, there are still some unresolved issues including the source of $CP$ violation~\cite{SM1}. In SM, $CP$ violation can be included by introducing a complex phase in the quark mixing matrix, which is named Cabibbo-Kobayashi-Maskawa~(CKM) matrix. Experimentally, starting in 1964, people subsequently observed $CP$ violation in the weak decay process of the $K$, $B$, and $D$ meson systems~\cite{KBmeson1,KBmeson2,KBmeson3,KBmeson4,KBmeson5,Dmeson}. The CKM quark mixing matrix can give a wonderful explanation of the observed $CP$ violation in the meson systems. However, the magnitude of $CP$ violation predicted by the SM cannot explain the matter-antimatter asymmetry in the universe~\cite{matter-antimatter asymmetry1,matter-antimatter asymmetry2,matter-antimatter asymmetry3}. Moreover, many extensions of the SM imply that the CKM matrix may not be the only source of $CP$ violation~\cite{SM extension1,SM extension2}. So more experimental studies are required to further test the $CP$ violation mechanism in SM and search for other sources of $CP$ violation. 
 
In 1956, \textit{Lee} and \textit{Yang} first proposed the violation of parity~($P$) conservation in the weak decays of baryons~\cite{LY}. The degree of violation can be expressed in terms of the asymmetry parameters, $\alpha = 2 Re~(s*p)/~(|s|^2+|p|^2)$, where $s$ and $p$ stand for the parity-violating $s$-wave and parity-conserving $p$-wave amplitudes in the weak decay. In 1986, theoretical physicist \textit{Pakvasa} proposed that the observable quantity of $CP$ violation could be constructed using asymmetric parameters in the decay of baryons, and predicted that the $CP$ violation of baryons in the SM is $\mathcal{O}~(10^{-5})$~\cite{Sandip Pakvasa, CPprediction}. 
The processes of pionic decays of hyperons provide a good place to explore $CP$ violation as they have a large branch ratio close to 1~\cite{hyperon1,hyperon2}. The $CP$ asymmetry can be described as $A_{CP}=\frac{\alpha+\bar{\alpha}}{\alpha-\bar{\alpha}}$, and the asymmetric parameters are $CP$-odd for the charge conjugate decay of $B/\bar{B}$~($B$ is a spin-1/2 baryon). Therefore, if $CP$ is conserved, $\alpha=-\bar{\alpha}$, $A_{CP}$ is equal to 0~\cite{hyperon1,hyperon2}. 

The Fermilab has specially designed HyperCP~(E871) experiment to study $CP$ violation of baryons in charged-$\Xi$ and $\Lambda$ hyperon decays. They have analyzed 11.7$\times10^{7}$ $\Xi^{-}\to \Lambda\pi^{-}\to p \pi^{-} \pi^{-}$ and 4.1$\times10^{7}$ $\Xi^{+}\to \Lambda\pi^{+}\to p \pi^{+} \pi^{+}$ events to determine the products $\alpha_{\Xi}\alpha_{\Lambda}$ and $\bar{\alpha}_{\Xi}\bar{\alpha}_{\Lambda}$~\cite{HyperCP}. The sum $A_{CP}^{\Lambda}$+$A_{CP}^{\Xi}$ was estimated to be~$(0.0\pm5.1\pm4.4)\times10^{-4}$~\cite{HyperCP}. In 2019, by studying the quantum entanglement of baryon pairs in the $J/\psi \to \Lambda \bar{\Lambda}$ process and using a multi-dimensional fitting method, the BESIII experiment obtained an independent measurement of $A_{CP}^{\Lambda}$ with matching precision: $A_{CP}^{\Lambda}=-0.006 \pm 0.012 \pm 0.007$, under the statistics of 0.4$\times10^{6}$ $J/\psi \to \Lambda \bar{\Lambda} \to p \pi^{-} \bar{p} \pi^{+}$ events~\cite{BESIII result}. Recently, the asymmetries from the direct and subsequent $J/\psi \to \Xi^{-}\bar{\Xi}^{+}$ decays were measured for the first time at BESIII and found to be $A_{CP}^{\Xi} = -0.0029\pm0.0133\pm0.0057$ and $\Delta\phi_{\Xi} = -0.0075\pm0.0137\pm0.0037$ rad~\cite{Xi}. Despite these, the $CP$ violation measurement accuracy of the current experiment still does not meet the prediction of the SM and is mainly dominated by statistics uncertainty~\cite{Sandip Pakvasa, CPprediction}.

To test for the existence of new sources of $CP$ violation other than SM, a hyperon sample with larger statistics is required. The STCF is a futural high-luminosity collider and also one of major options for the accelerator-based high-energy project in China in the post-BEPCII era. The center-of-mass energy~$(\sqrt{s})$ of the STCF collision will cover $2 \sim 7$~GeV, which has been doubled compared to BEPCII. The peaking luminosity is expected to be over $0.5\times10^{35}$~cm$^{-2}$~s$^{-1}$ or higher at $\sqrt{s}=4$~GeV. It is expected to provide more than $1.0\times10^{12}$ $J/\psi$ events per year and has great potential for improving luminosity and realizing beam polarization. So STCF will be an ideal place to study $CP$ violation of $\Lambda$ decay. 

In this analysis, we performed the sensitivity study of decay asymmetries of $\Lambda$ decay and the decay channel is $e^{+}e^{-}\to J/\psi\to \Lambda~(\to p\pi^{-})\bar{\Lambda}~(\to \bar{n}\pi^{0})$ with the statistics of $1.0\times10^{12} J/\psi$ MC events. The amplitude of the signal process follows the helicity amplitude method which is described explicitly, as shown in Eq.~\ref{eqw}.  
Furthermore, the final states of $\Lambda \to p \pi^{-}$ decay have one low-momentum $\pi^{-}$ particle, which plays a key role in limiting the overall reconstruction efficiency. Therefore, it is essential to improve the reconstruction efficiency of the low-momentum $\pi^{-}$ to get better sensitivity, so the decay of $J/\psi \to \Lambda \bar{\Lambda}$ is also used as a benchmark process in this analysis to perform optimization of detector performance design.


\section{Formalism}
\label{formalism}
\begin{figure}[hbpt]
        \begin{center}
            \includegraphics[width=0.5\textwidth,height=0.18\textheight]{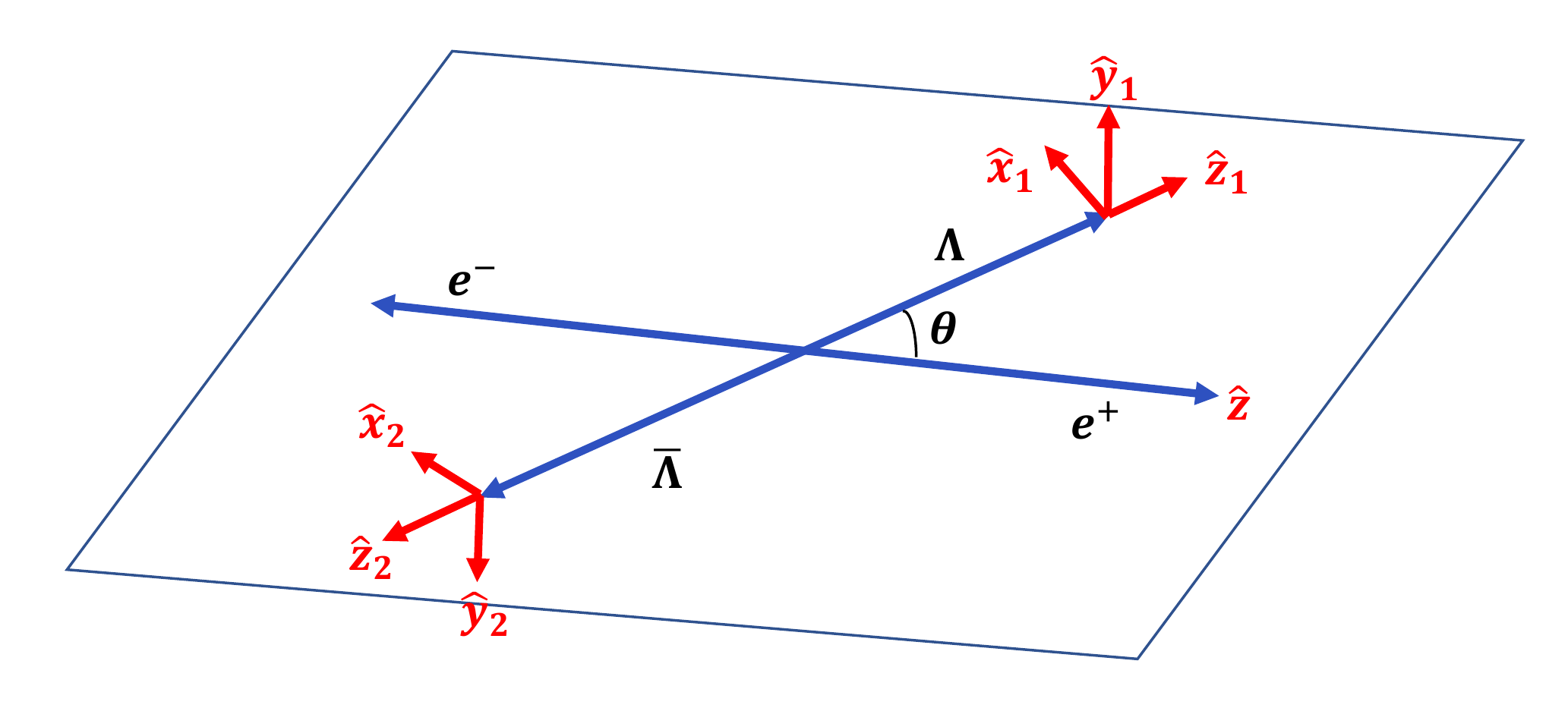}
            \caption{The reaction system with the defined helicity angles in $\Lambda\bar{\Lambda}$ decay.}
            \label{decay_coordinate}
        \end{center}
    \end{figure}

The production process $e^{+}e^{-}\to J/\psi\to\Lambda\bar{\Lambda}$ is described in the c.m. system of $J/\psi$. The scattering angle $\theta$ of $\Lambda$ is defined by
\begin{equation}
\cos\theta =\boldsymbol{ \hat{p}\cdot\hat{k}}, 
\end{equation}
where $\boldsymbol{p}$ and $\boldsymbol{k}$ are the three momenta of outgoing $\Lambda$ and initial positron, respectively. The scattering plane with the vector $\boldsymbol{p}$ and $\boldsymbol{k}$ is used to form the $xz$-plane, and the corresponding $y$-axis is perpendicular to the scattering plane. The right-handed coordinate system is defined as follows:
\begin{equation}
\begin{aligned}
& \boldsymbol{e_{x}=\frac{1}{\sin\theta}(\hat{k}\times\hat{p})\times\hat{p},}\\
& \boldsymbol{e_{y}=\frac{1}{\sin\theta}(\hat{k}\times \hat{p}),}\\
& \boldsymbol{e_{z}=\hat{p}.} 
\end{aligned}
\end{equation}
The spin density matrix for a two spin 1/2 particle state can be expressed in terms of a set of 4 $\times$ 4 matrices obtained from the outer product, $\otimes$, of $\sigma_{\mu}$ and $\sigma_{\bar{\nu}}$~\cite{density matrix}:
\begin{equation}
\begin{aligned}
& {\rho=\cfrac{1}{4}\sum_{{\mu}\bar{\nu}}C_{{\mu}\bar{\nu}}\sigma_{\mu}^{\Lambda}\otimes\sigma_{\bar{\nu}}^{\bar{\Lambda}}}, 
\end{aligned}
\end{equation}
where $\sigma_{\mu,\bar{\nu}}$ with $\mu,\bar{\nu}$= 0, 1, 2, 3, represent spin-$1/2$ base matrices for baryon $\Lambda/\bar{\Lambda}$ in the rest frame. The 2 × 2 matrices are $\sigma_{0} =1_{2}$, $\sigma_{1}=\sigma_{x}$, $\sigma_{2}=\sigma_{y}$, and $\sigma_{3}=\sigma_{z}$. In particular, the spin matrices $\sigma_{\mu}$ and $\sigma_{\bar{\nu}}$ are given in the helicity frames of the baryons $\Lambda$ and $\bar{\Lambda}$, respectively. We define the coordinate system for $\Lambda\bar{\Lambda}$ decay, as shown in Fig.~\ref{decay_coordinate}. The real coefficients $C_{\mu\bar{\nu}}$ for $e^{+}e^{-} \to J/\psi \to \Lambda \bar{\Lambda}$ with non-polarized inject beams are given by Eq.~\ref{eq4}, 
\begin{equation}
\begin{aligned}
\label{eq4}
&C_{\mu\bar{\nu}}=\\
&\begin{pmatrix}
1+\alpha\cos^2\theta       & 0                          & \beta\sin\theta\cos\theta& 0 \\
0                          & \sin^2\theta                & 0                        & \gamma\sin\theta\cos\theta \\
-\beta\sin\theta\cos\theta & 0                          &\alpha\sin^2\theta        & 0 \\
0                          & -\gamma\sin\theta\cos\theta&0                         & -\alpha-\cos^2\theta 
\end{pmatrix}
\end{aligned}{}
\end{equation}
, where $\beta=\sqrt{~(1-\alpha^2)}\sin~(\Delta\Phi)$ and $\gamma=\sqrt{~(1-\alpha^2)}\cos~(\Delta\Phi)$, are functions of the scattering angle $\theta$ of $\Lambda$. In the real coefficients $C_{\mu\bar{\nu}}$ of Eq.~\ref{eq4}, there are two parameters related to the production process of $e^{+}e^{-}\to J/\psi\to\Lambda\bar{\Lambda}$, the ratio of two helicity amplitudes $\alpha$, and the relative phase of the two helicity amplitudes $\Delta\Phi$.

After considering the subsequent two-body weak decays into $p\pi^{-}$/$\bar{n}\pi^{0}$, the joint angular distribution of the $p/\bar{n}$ pair is given within the present formalism as~\cite{density matrix}:

\begin{equation}
\begin{aligned}
\label{eq5}
& {Tr\rho_{p\bar{n}}\propto\sum_{\mu,\bar{\nu}=0}^3 C_{\mu\bar{\nu}}~(\theta)a^{\Lambda}_{\mu0}a^{\bar{\Lambda}}_{\bar{\nu}0}}, 
\end{aligned}
\end{equation}
where the $a_{\mu0}^{\Lambda}~(\theta_{1},\phi_{1};\alpha_{1})$ and $a_{\bar{\nu}0}^{\bar{\Lambda}}~(\theta_{2},\phi_{2};\alpha_{2})$ represent the correlation of the spin density matrices in the sequential decays and the full expressions can be found in Ref.~\cite{density matrix}. $\alpha_{1}/\alpha_{2}$ are the decay asymmetries for $\Lambda \to p \pi^{-}/\bar{\Lambda} \to \bar{n} \pi^{0}$. The variables $\theta_{1}$ and $\phi_{1}$ are the proton spherical coordinates in the $\Lambda$ helicity frame with the axes $\boldsymbol{x_{1}}, \boldsymbol{y_{1}}, \boldsymbol{z_{1}}$ defined in Fig.~\ref{decay_coordinate}. The variables $\theta_{2}$ and $\phi_{2}$ are the anti-neutron spherical angles in the $\bar{\Lambda}$ helicity frame with the axes $\boldsymbol{x_{2}}, \boldsymbol{y_{2}}, \boldsymbol{z_{2}}$. 

An event of the reaction $e^{+}e^{-}\to J/\psi\to \Lambda~(\to p\pi^{-})\bar{\Lambda}~(\to \bar{n}\pi^{0})$ is specified by the five-dimensional vector $\xi=~(\theta,\Omega_{1}~(\theta_{1},\phi_{1}), \Omega_{2}~(\theta_{2},\phi_{2}))$, and the joint angular distribution $\mathcal{W~(\xi)}$ can be expressed as:
\begin{equation}
\begin{aligned}
\mathcal{W~(\xi)} & = \mathcal{F}_{0}~(\xi) + \alpha \mathcal{F}_{5}~(\xi)\\
& + \alpha_{1}\alpha_{2}~(\mathcal{F}_{1}~(\xi) + \sqrt{1-\alpha^{2}}\cos~(\Delta\Phi)\mathcal{F}_{2}~(\xi) +\alpha\mathcal{F}_{6}~(\xi))\\
& + \sqrt{1-\alpha^{2}}\sin~(\Delta\Phi)~(-\alpha_{1}\mathcal{F}_{3}~(\xi) + \alpha_{2} \mathcal{F}_{4}~(\xi))
\end{aligned}
\label{eqw}
\end{equation}
with a set of angular functions $\mathcal{F}_{i}~(\xi)$ defined as:
\begin{equation}
\begin{aligned}
&\mathcal{F}_{0}~(\xi)=1 \\
&\mathcal{F}_{1}~(\xi)= \sin^{2}\theta\sin\theta_{1}\sin\theta_{2}\cos\phi_{1}\cos\phi_{2} - \cos^{2}\theta\cos\theta_{1}\cos\theta_{2}  \\
&\mathcal{F}_{2}~(\xi)= \sin\theta\cos\theta~(\sin\theta_{1}\cos\theta_{2}\cos\phi_{1} - \cos\theta_{1}\sin\theta_{2}\cos\phi_{2}) \\
&\mathcal{F}_{3}~(\xi)= \sin\theta\cos\theta\sin\theta_{1}\sin\phi_{1} \\
&\mathcal{F}_{4}~(\xi)=\sin\theta\cos\theta\sin\theta_{2}\sin\phi_{2} \\
&\mathcal{F}_{5}~(\xi)= \cos^{2}\theta \\
&\mathcal{F}_{6}~(\xi)= \sin^{2}\theta\sin\theta_{1}\sin\theta_{2}\sin\phi_{1}\sin\phi_{2}-\cos\theta_{1}\cos\theta_{2}. 
\end{aligned}
\end{equation}

There are four terms in Eq.~\ref{eqw}: the first two~$(\mathcal{F}_{0}+\alpha\mathcal{F}_{5})$ describe the production angular distribution, and the third and fourth terms give the spin correlation and polarization, respectively. The polarization is in the $\boldsymbol{e_{y}}$ direction and is related to the phase $\Delta\Phi$ via~\cite{polar}
\begin{equation}
\label{pol}
P_{y}=-\frac{\sqrt{1-\alpha^{2}}\sin\theta\cos\theta}{1+\alpha\cos^{2}\theta}\sin~(\Delta\Phi).
\end{equation}
The polarization can only occur when $\Delta\Phi$ is not equal to 0. As a consequence, the decay asymmetries can be determined with nonzero $\Delta\Phi$. Using this conclusion, the BESIII experiment used the angular distribution analysis method to observe the nonzero relative phase $\Delta\Phi$ of $\Lambda$ in the baryon system for the first time, and then measured the decay asymmetry of $\Lambda$ decay~\cite{BESIII result}.

\section{Detector and MC simulations}
The design structure of the STCF detector from the interaction point to the outside mainly includes a tracking system, a particle identification~(PID) system, an electromagnetic calorimeter~(EMC), a super-conducting solenoid and a muon detector~(MUD). The detailed conceptual design of each sub-detector can be found in~\cite{Fastsim1,Fastsim2}. 

The STCF detector and offline software system are under research and development at present. In order to study the physical potential of STCF and further optimize the detector design, a fast simulation software package dedicated to STCF detectors has been developed~\cite{Fastsim1,Fastsim2} and it has proven to be a useful tool for analysis in STCF. The fast simulation is simple to use and can simulate the response of objects in each sub-detector without ${\sc Geant4}$, including variables such as efficiency, and resolution~(space, momentum, energy, time, etc.). 
By default, all the parameterized parameters for each sub-detector performance are based on the BESIII performance~\cite{BESIII}, but can be adjusted flexibly by scaling a factor according to the expected performance of the STCF detector, or by implementing a special interface to model any performance described with an external histogram, an input curve, or a series of discrete data~\cite{Fastsim1}. In this analysis, the default scale factor is set to 1.0, which can be used to optimize the detector design according to physical requirements. 
 
\section{Analysis of $J/\psi\to\Lambda\bar{\Lambda}$ with fast simulation}

The $J/\psi\to\Lambda\bar{\Lambda}$ reaction is identified with the $\Lambda$ subsequently decaying into $p\pi^{-}$ and $\bar{\Lambda}$ decay into $\bar{n}\pi^{0}$ resulting in a final state of $p\pi^{-}\bar{n}\gamma\gamma$. So, the candidate events are required to have at least two oppositely charged tracks and at least three showers. 

The combination of positive and negative charged tracks closest to the PDG mass of $\Lambda$ was chosen as the $\Lambda$ candidate~\cite{PDG}. In addition, the two daughter tracks are constrained to originate from a common decay vertex. The most energetic shower with energy deposition greater than 350~MeV is selected as $\bar{n}$. The two showers except the $\bar{n}$ candidate are consistent with photons and are used to reconstruct the $\pi^{0}$ candidates. At least, one good $\pi^{0}$ is required. In order to select the $J/\psi\to\Lambda~(p\pi^{-})\bar{\Lambda}~(\bar{n}\pi^{0})$ candidate events, a two-constrained~(2C) kinematic fit was performed, where $\bar{n}$ is treated as a missed particle with mass fixed to 0.938~GeV~\cite{PDG}, and the constraints including the four-momentum conservation of $J/\psi$ and an additional constraint of photon pair to have an invariant mass equal to $\pi^{0}$. Furthermore, $\theta_{\bar{n}}$ is required to be less than $5^{\circ}$, where $\theta_{\bar{n}}$ is defined as the angle between the $\bar{n}$ direction obtained from kinematic fit and the most energetic shower. To further suppress the background, $\Lambda$ and $\bar{\Lambda}$ candidates are required to be within 1.110~GeV/$c^{2} < M_{p\pi^{-}} < 1.120$~GeV/$c^{2}$ and 1.098~GeV/$c^{2} < M_{\bar{n}\pi^{0}} < 1.127$~GeV/$c^{2}$.

The $1.0\times10^{6}$ events of the $J/\psi\to\Lambda\bar{\Lambda}\to p\pi^{-}\bar{n}\pi^{0}$ process were generated to optimize the selection criteria and evaluate the selection efficiencies for the baryon pair production. Based on the above selection conditions, with the help of fast simulation software, 129575 candidate events of $J/\psi\to\Lambda\bar{\Lambda}\to p\pi^{-}\bar{n}\pi^{0}$ were selected. The step-by-step selection efficiency is shown in Table~\ref{effloss}.

Furthermore, these MC samples also are used to optimize the detector response and $1.0\times10^{12}$ events of signal process were generated to test the sensitivity of $CP$ violation. To analyze the potential background process, $1.0\times10^{6}$ events of $J/\psi \to anything$ were generated as the inclusive MC. After the above event selection criteria were applied on the inclusive MC and by topology analysis, the $J/\psi\to\Lambda\bar{\Sigma}^{0}\to p\pi^{-}\bar{n}\pi^{0}\gamma$ process has be shown to be the dominant background. So, $1.0\times10^{12}$ events of the $J/\psi\to\Lambda\bar{\Sigma}^{0}\to p\pi^{-}\bar{n}\pi^{0}\gamma$ process were generated to do the background test in the next chapter. Furthermore, 0.7$\times10^{9}$ events of the signal process were generated using the phase space~(PHSP) generator to estimate the normalization coefficient in Maximum Likelihood~(MLL) fit.

\section{Optimization of detector performance}

After the above event selection, the final selection efficiency is about $12.96\%$. The performance of the detector can be optimized from the following aspects: the selection efficiency of the charged tracks, the momentum resolution of the charged tracks, and the position resolution of the photons. Utilizing the signal MC sample and with the help of fast simulation software tools, the optimized results of the detector response are as follows:
 
~{\it a.Tracking efficiency}

The charged particles in the final state that can be identified by the detector include electrons, muons, pions, kaons, and protons. These charged particles have a wide range of momentum, some can be as high as 3.5~GeV/c, and some can be less than 1~GeV/c. This situation requires the detector to have the ability to cover a large momentum range and high-reconstruction efficiency. In the part of track system design of STCF, different materials or advanced tracking algorithms can be used to further improve the ability of low-momentum track reconstruction. The $J/\psi\to\Lambda\bar{\Lambda}\to p\pi^{-}\bar{n}\pi^{0}$ decay has low-momentum final state particle $\pi^{-}$, which is a good choice for optimizing the detector response, improving the resolution of low-momentum particles.

\begin{figure}[hbpt]{}
    \begin{center} 
    {
        \centering
        \includegraphics[width=0.5\textwidth,height=0.28\textheight]{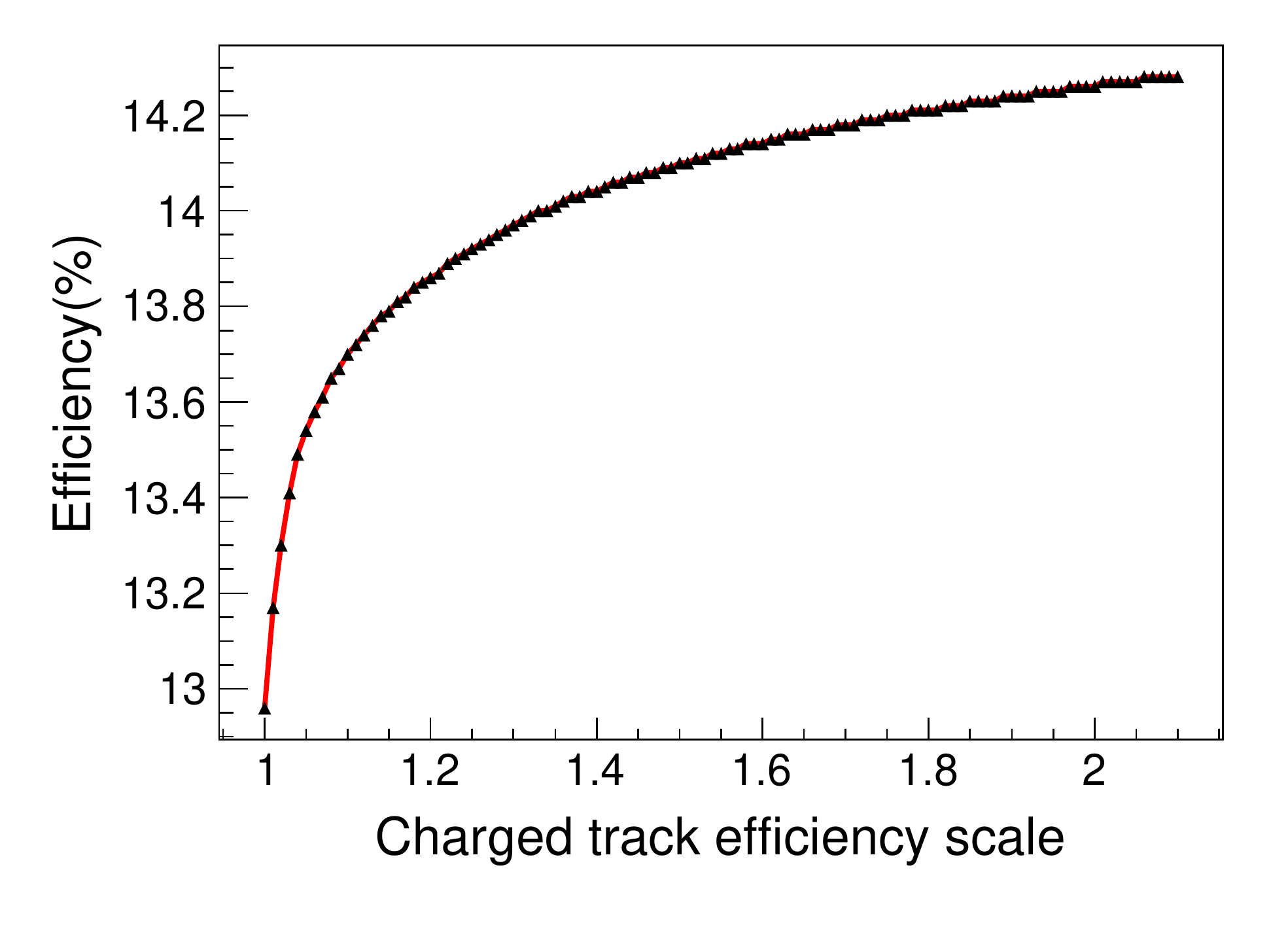}
    }            
    \caption{Charged track efficiency scale versus the selection efficiency.}
    \label{chargeeff}
    \end{center}
\end{figure}

In this analysis, we gradually adjusted the scale factor of tracking efficiency from 1.0 to 2.0. It can be seen from Fig.~\ref{chargeeff} that the final selection efficiency has increased significantly in the range from 1.0 to 1.1 of the scale factor, and the selection efficiency will increase from 12.96$\%$ to 13.67$\%$.

~{\it b.Momentum resolution of the charged tracks}

\begin{figure}[hbpt]{}
    \begin{center} 
    {
        \centering
        \includegraphics[width=0.5\textwidth,height=0.28\textheight]{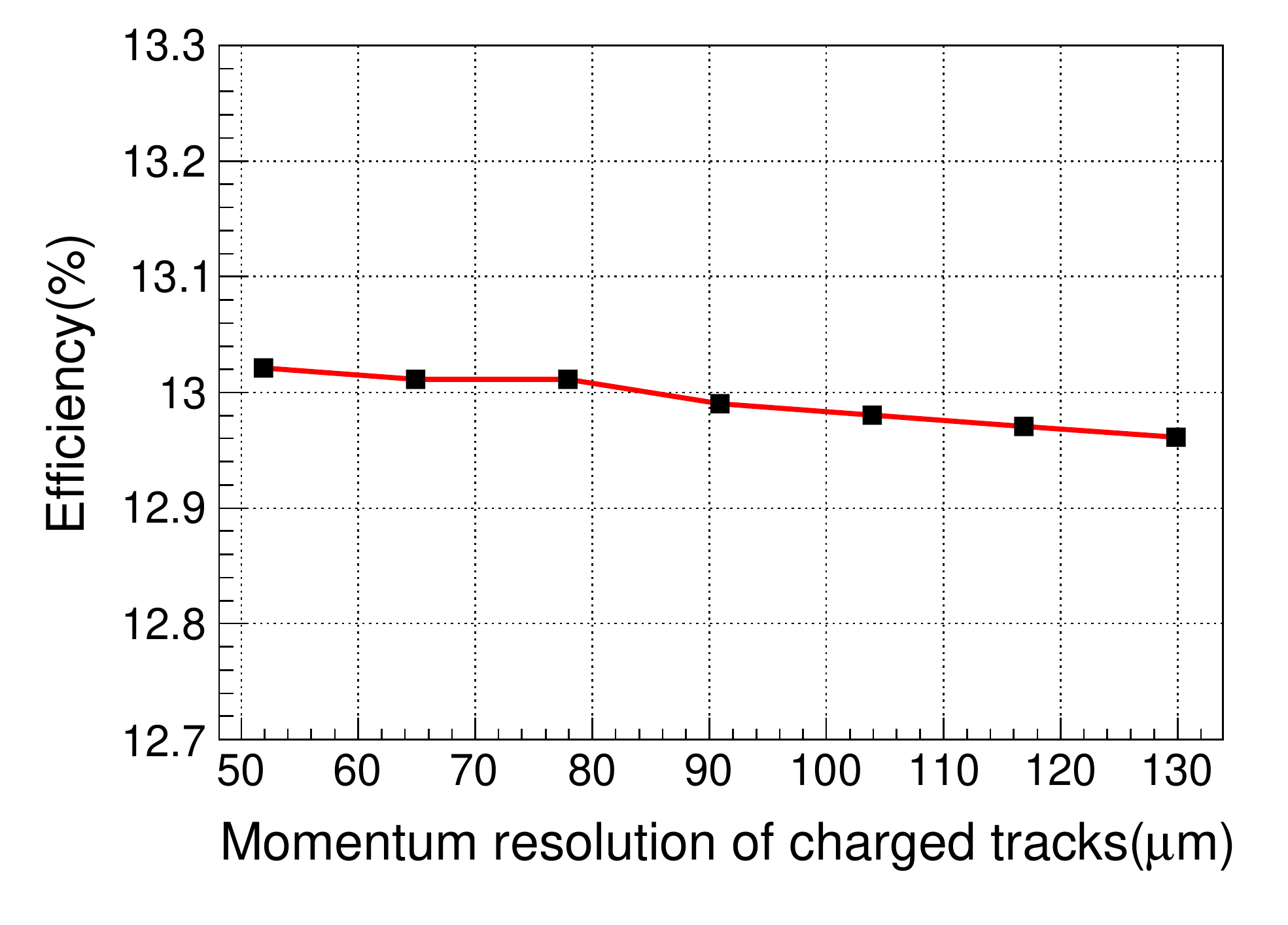}
    }            
    \caption{Momentum resolution of charged tracks versus the selection efficiency.}
    \label{chargeXYZ}
    \end{center}
\end{figure}

The momentum resolution of the charged tracks can also be optimized by the fast simulation. $\sigma_{xy}$ and $\sigma_{z}$ are the spatial resolutions of tracks in the $xy$-plane and $z$-direction. By default, $\sigma_{xy}=130$~$\mu$m and $\sigma_{z}$=2480~$\mu$m. Optimizing $\sigma_{xy}$ from 52~$\mu$m to 130~$\mu$m, and the corresponding $\sigma_{z}$ is optimized from 992~$\mu$m to 2480~$\mu$m. There is no significant change in efficiency, as shown in Fig.~\ref{chargeXYZ}. 

\begin{figure}[hbpt]{}
    \begin{center} 
    {
        \centering
        \includegraphics[width=0.5\textwidth,height=0.28\textheight]{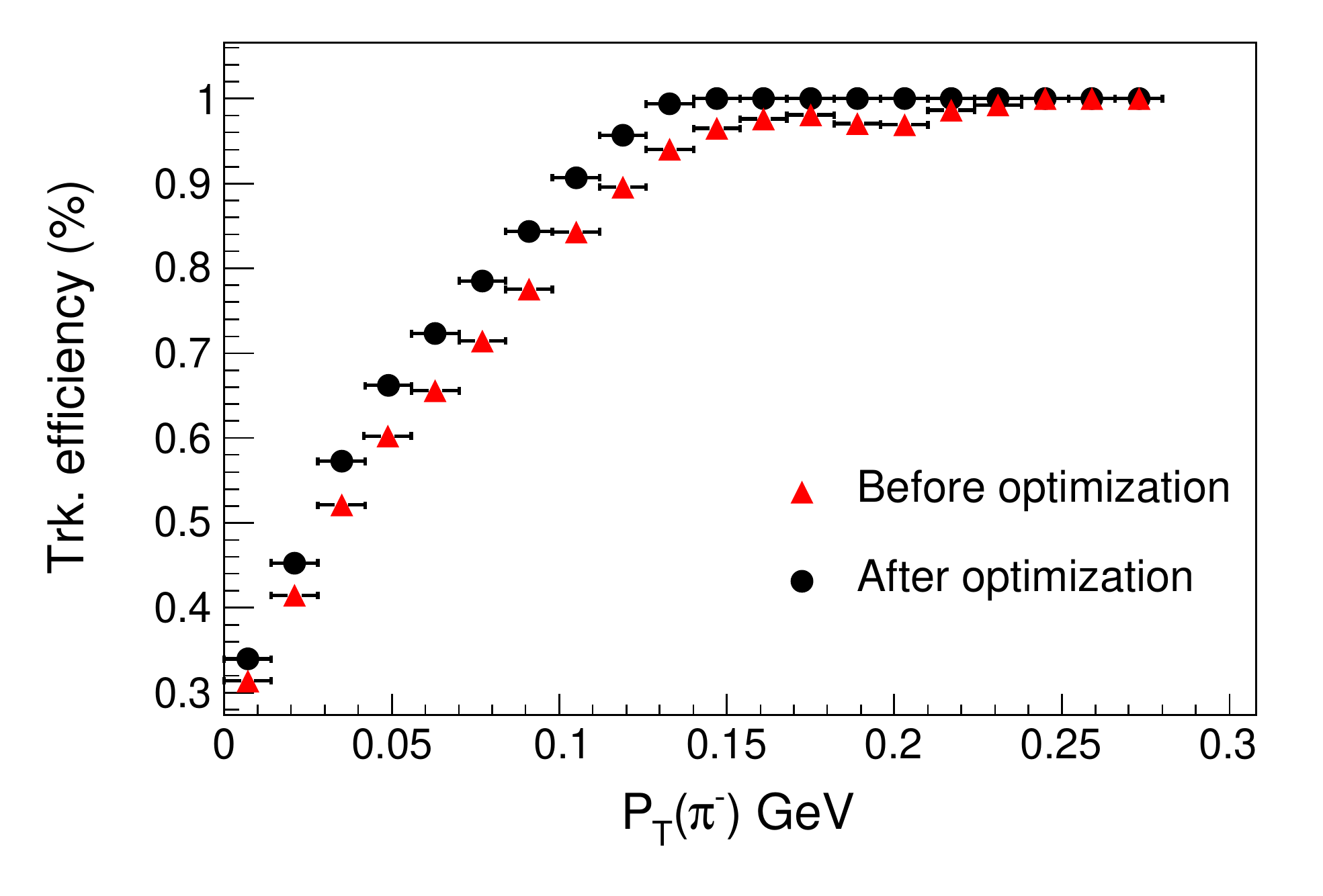}
    }            
    \caption{The Optimization curve of the transverse momentum of $\pi$.}
    \label{pt}
    \end{center}
\end{figure}

In addition, the transverse momentum $P_{T}$ and polar angle $\cos\theta$ are two characteristic quantities of track reconstruction in MDC. They are related to the level of track bending and hit positions of tracks in the MDC. The optimization curve of the transverse momentum of low-momentum $\pi^{-}$ is shown in Fig.~\ref{pt}, where the black and red points represent the ratio of signal efficiency to MC truth before and after all the above optimization, respectively.

~{\it c.Position resolution of photon}

\begin{figure}[hbpt]{}
    \begin{center} 
    {
        \centering
        \includegraphics[width=0.5\textwidth,height=0.28\textheight]{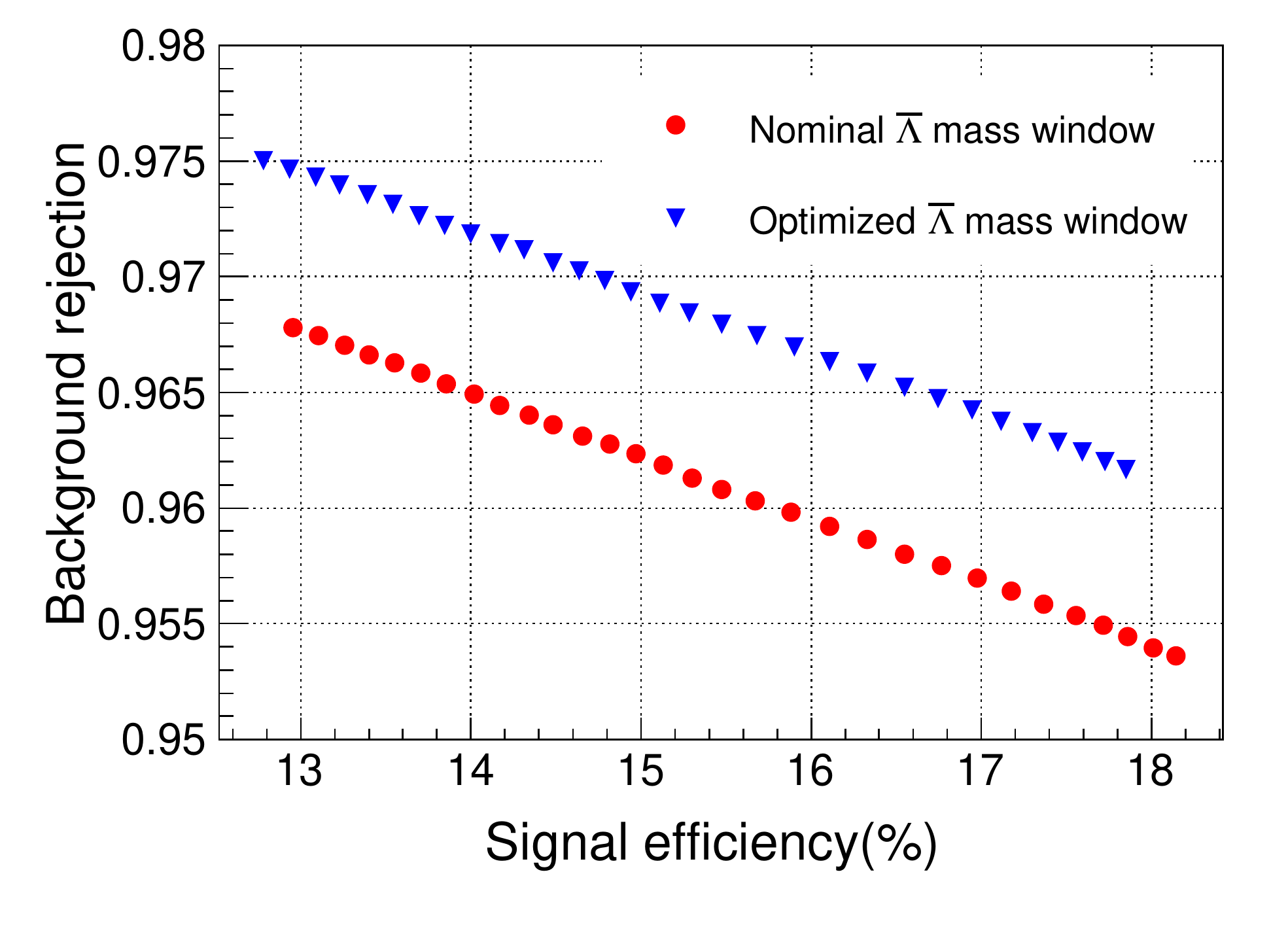}
    }            
    \caption{The change of signal selection eﬀiciency and background rejection with position resolution of the photon.}
    \label{backgroundrejection}
    \end{center}
\end{figure}

The decay of $J/\psi\to\Lambda\bar{\Lambda}\to p\pi^{-}\bar{n}\pi^{0}$ has a final state particle $\pi^{0}$, $\pi^{0}$ is reconstructed by two photons, so this process is also very sensitive to the EMC performance. With the increase in the resolution of the $\pi^{0}$, there will be a better signal-to-background ratio and higher detection efficiency. Optimizing the signal-to-background ratio can provide a reference for the EMC design. In this analysis, the signal process $J/\psi\to\Lambda\bar{\Lambda}\to p\pi^{-}\bar{n}\pi^{0}$ and the main background process $J/\psi\to\Lambda\bar{\Sigma}^{0}\to p\pi^{-}\bar{n}\pi^{0}\gamma$ were studied. By fitting the distribution of the invariant mass of M$_{\bar{n}\pi^{0}}$, a $3\sigma$ mass interval of M$_{\bar{n}\pi^{0}}$ is obtained to further reduce the impact of the background process. Figure~\ref{backgroundrejection} shows the signal selection efficiency and background rejection under the change of photon position resolution. The scale factor of the position resolution of photon varies from 0.4 to 1.0. The red and blue points represent the case of using the nominal $\bar{\Lambda}$ mass window and the optimized $\bar{\Lambda}$ mass window, respectively. Although this will lose some signal events, it can reduce more background and make the signal cleaner. It is appropriate to set the scale factor to 0.7, which corresponds to the position resolution of 4~mm. The signal selection efficiency will get increase from $12.96\%$ to $15.11\%$, while the main background will be reduced from $3.27\%$ to $3.17\%$.

After all the optimization of detector responds, the events of signal MC will increase from $12.96\%$ to $15.97\%$, while the events of main background~($J/\psi\to\Lambda\bar{\Sigma}^{0}\to p\pi^{-}\bar{n}\pi^{0}\gamma$) will reduce from $3.27\%$ to $3.09\%$. The selection efficiency is as shown in Table~\ref{effloss}. 
\begin{table*}[ht]
\centering
\footnotesize
\caption{Events selection efficiency.}
\label{effloss}
\begin{tabular}{c c c c c c}
\hline\hline
\multirow{2}{*}{\textbf{  } }&\multirow{2}{*}{\textbf{No optimized eff.~(\%)}}&\multicolumn{1}{c}{\multirow{2}{*}{\textbf{Optimized eff.~(\%)}}}&\textbf{Increased efficiency after}\\
&&&\textbf{optimization in step~(\%)}\\
\hline
Charged \ tracks                       			 &  74.21  &  79.38  & 5.17		\\[6pt]  
$\Lambda$\ reconstruction                      	 &  66.27  &  70.88  & 4.61		\\[6pt]  
Good\ showers                                  	 &  31.73  &  33.57  & 1.84		\\[6pt]  
$\pi^{0}$\ 1C \ fit\ (N$_\gamma\ge$2)            &  28.76  &  29.93  & 1.17		\\[6pt]  
Kinematic \ 2-C \ fit                          	 &  25.33  &  27.31  & 1.98		\\[6pt] 
Energy \ deposition\ of\ $\bar{n}>$0.35~GeV      & 21.18   &  22.88  & 1.70		\\[6pt] 
$\theta_{\bar{n}}<5^{\circ}$               	 &  14.54  &  18.34  & 3.80		\\[6pt] 
$\Lambda$ and $\bar{\Lambda}$\ mass \ window     &  12.96  &  15.97  & 3.01		\\[6pt]
\hline
\hline
\end{tabular}
\end{table*}

\section{Extraction of the parameters}

In this analysis, the parameters can be extracted by applying an unbinned MLL fit. The probability density function of the $i$th event can be expressed by
\begin{equation}
\mathcal{P}~(\xi_{i};pars) = \mathcal{W}~(\xi_{i};pars)\epsilon~(\xi_{i})/\mathcal{N}~(pars)
\end{equation}
, where $\epsilon~(\xi_{i})$ is the efficiency of each event, $\xi_{i}$ and $pars$ are a set of angular vectors and parameters: $\xi_{i}=~(\theta,\Omega_{1},\Omega_{2})$, $pars$=$~(\alpha, 
\alpha_{1}, \alpha_{2}, \Delta\Phi)$, as described in Sec.~\ref{formalism}.

\begin{figure}[hbpt]{}
    \begin{center} 
    {
        \centering
        \includegraphics[width=0.5\textwidth,height=0.28\textheight]{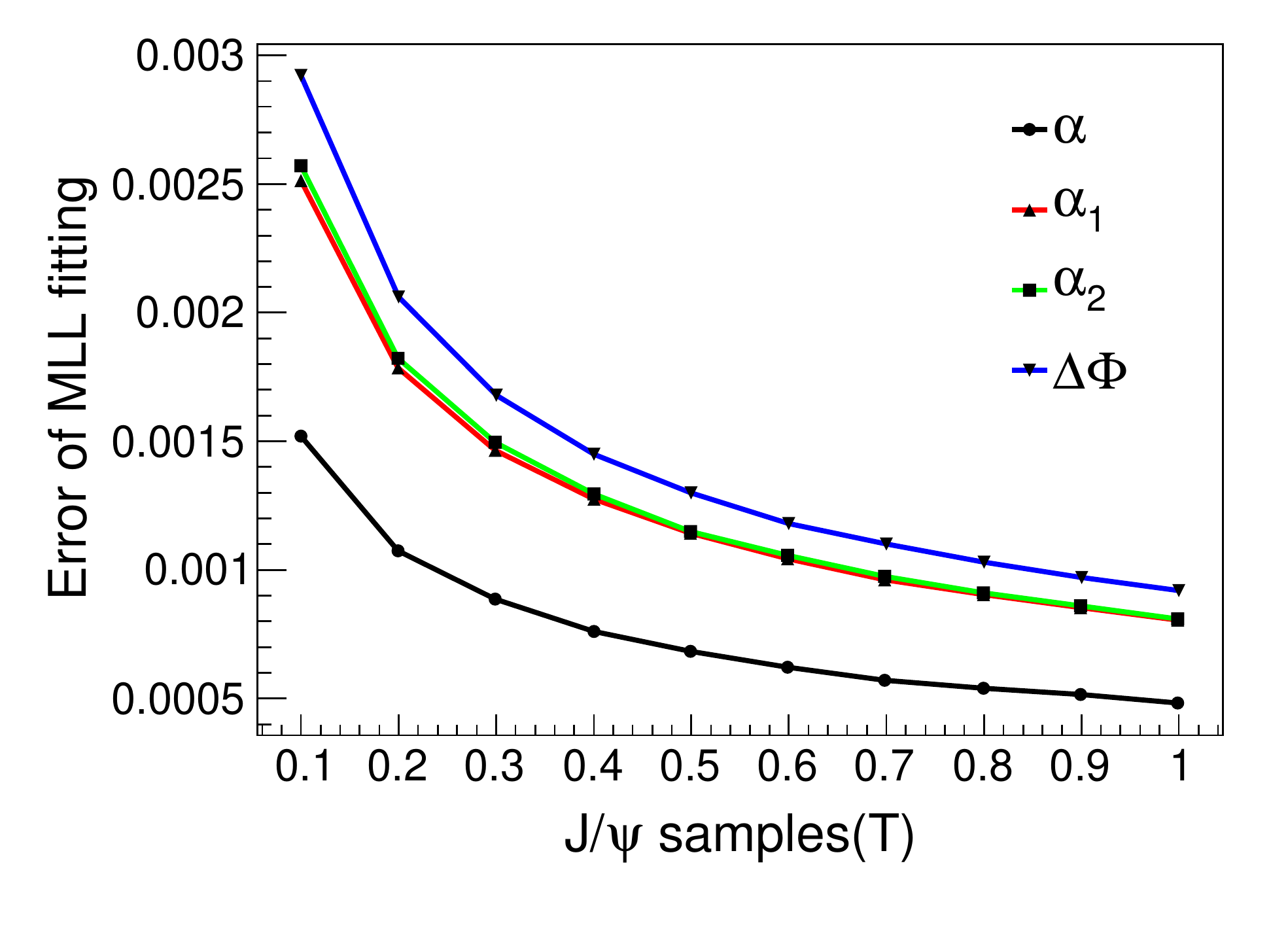}
    }            
    \caption{The statistical sensitivity of $J/\psi$ samples with different statistics.}
    \label{sensitivity}
    \end{center}
\end{figure}

The joint probability density for observing $N$ events in the data sample is~\cite{propability density}:
\begin{equation}
\begin{aligned}
 \mathcal{P}~(\xi_{1},\xi_{2},...,\xi_{N};pars) =\prod_{i=1}^{N}\mathcal{P}~(\xi_{i};pars) \\
=\prod_{i=1}^{N}\frac{\mathcal{W}~(\xi_{i};pars)\epsilon~(\xi_{i})}{\mathcal{N}~(pars)}.
\end{aligned}
\end{equation}
By taking the natural logarithm of the joint probability density, the efficiency function can be separated
\begin{equation}
{\rm \ln}\mathcal{P}~(\xi_{1}...,\xi_{N};pars)=\sum_{i}^{N}{\rm ln}\frac{\mathcal{W}~(\xi_{i};pars)}{\mathcal{N}~(pars)} +\sum_{i}^{N}{\rm ln}\epsilon~(\xi_{i}).
\end{equation}
Usually, the minimization of -ln$\mathcal{L}$ is performed by using {\sc MINUIT}~\cite{Minuit} 
\begin{equation}
-{\rm ln}\mathcal{L}=-\sum_{i}^{N}{\rm ln}\frac{\mathcal{W}~(\xi_{i};pars)\epsilon~(\xi_{i})}{\mathcal{N}~(pars)}
\end{equation}
, where $\mathcal{N}$ is the normalization factor, given by
\begin{equation}
\mathcal{N}=\int \mathcal{W}~(\xi)\epsilon~(\xi_{i})d\cos\theta d\Omega_{1} d\Omega_{2}.
\end{equation}
For a certain set of pars, $\mathcal{N}$~(pars) can be rewritten as the integration on each $\mathcal{F}_{i}$ term according to Eq.~\ref{eqw}. To test the statistical sensitivity, the fitting was applied on $J/\psi$ samples with different statistics. The precision for the decay parameters is shown in Fig.~\ref{sensitivity}. It is found that the precision of the parameters is proportional to the square root of the $J/\psi$ sample. The correlation matrix among the parameters is shown in Table~\ref{corre}.

\begin{table}[htbp!]
\centering
\caption{Correlation Matrix for the parameters, obtained with {\sc MINUIT}.}
\label{corre}
\begin{tabular} {ccccc }
\hline\hline
$pars$      &  $\alpha$  & $\alpha_{1}$ & $\alpha_{2}$ & $\Delta\Phi$ \\
$\alpha$    &  1.000     & -0.089       & 0.104        & 0.339\\
$\alpha_{1}$&  -0.089    & 1.000        & 0.853        & -0.120\\
$\alpha_{2}$&  0.104     & 0.853        & 1.000        & 0.058 \\
$\Delta\Phi$&  0.339     & -0.120       & 0.058        & 1.000\\
\hline\hline
\end{tabular}
\end{table}

According to Eq.~\ref{eqw}, the moment of $\sin\theta_{1}\sin\phi_{1}$ is given by
\begin{equation}
\begin{aligned}
\langle\sin\theta_{1}\sin\phi_{1}\rangle& = \frac{1}{N_{norm}} \int \mathcal{W}~(\xi)\sin\theta_{1}\sin\phi_{1} d\Omega_{1}d\Omega_{2}\\
& \approx -\frac{\sqrt{1-\alpha^{2}}\alpha_{1}\sin~(\Delta\Phi)}{3+\alpha}\sin\theta\cos\theta.
\end{aligned}
\end{equation}

In the analysis of experimental data, $\langle\sin\theta_{1}\sin\phi_{1}\rangle$ can be calculated by the average of $\sin\theta_{1}\sin\phi_{1}$ in each $\cos\theta$ bin.
%
The moment of $\sin\theta_{1}\sin\phi_{1}$ can be connected with the polarization according to Eq.~\ref{pol},
\begin{equation}
\langle\sin\theta_{1}\sin\phi_{1}\rangle \approx \frac{~(1+\alpha\cos^{2}\theta)\alpha_{1}}{3+\alpha}P_{y}.
\end{equation}

\begin{figure}[hbpt]{}
    \begin{center} 
    {
        \centering
        \includegraphics[width=0.5\textwidth,height=0.28\textheight]{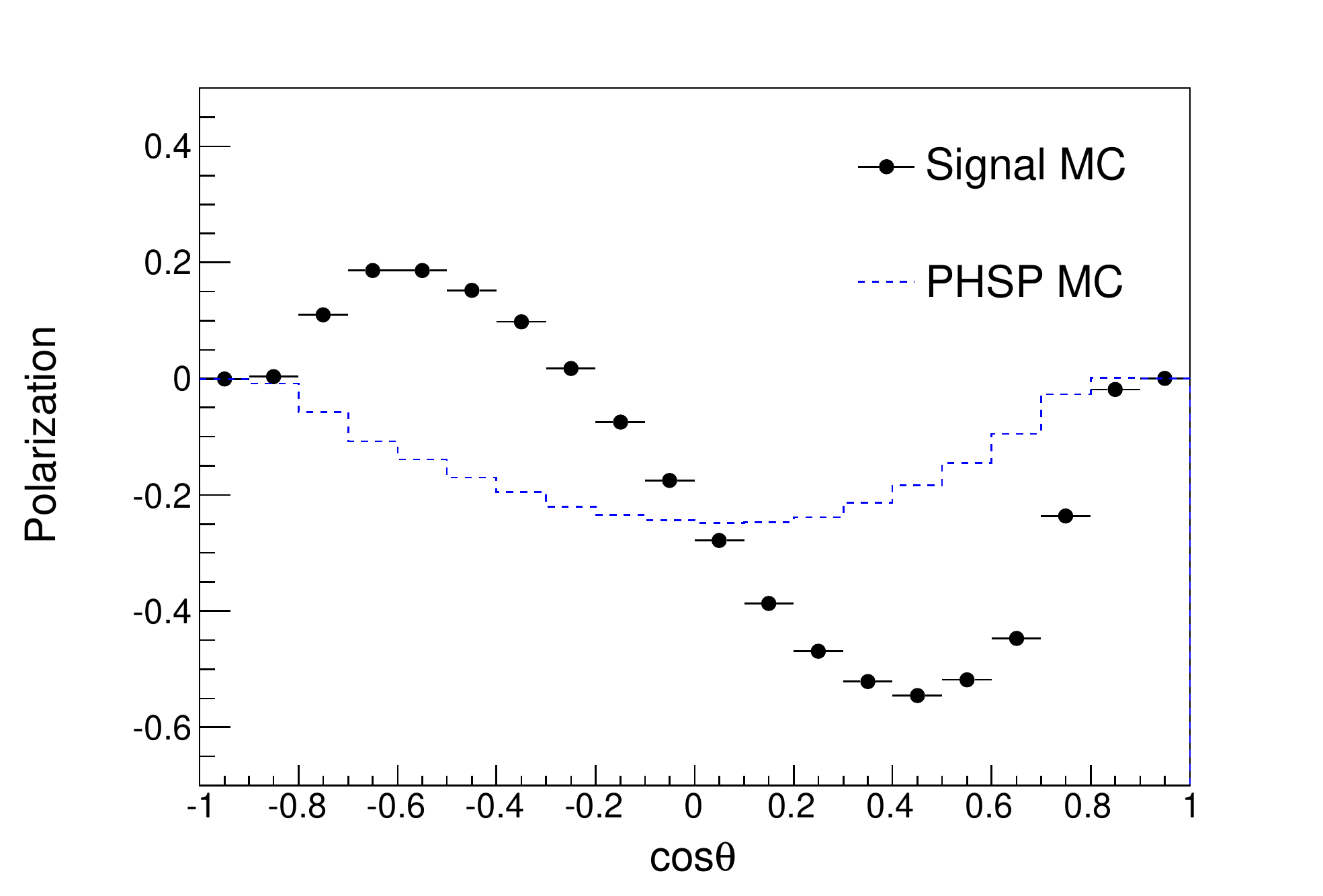}
    }            
    \caption{Polarization as a function of $\cos\theta$ for $J/\psi \to \Lambda \bar{\Lambda} \to p\pi^{-}\bar{n}\pi^{0}$. The points with error bars are the signal MC, and the blue dashed histogram is the no-polarization scenario of PHSP MC.}
    \label{polar}
    \end{center}
\end{figure}

The distribution of polarization versus $\cos\theta$ as shown in Fig.~\ref{polar} and the events are not corrected with detection efficiency.

\section{ Prospect of $CP$ sensitivity at STCF }

The asymmetry parameters used to observe $CP$ violation are affected by statistics and proportional to the $\sqrt{N_{J/\psi}}$, where $N_{J/\psi}$ is the number of $J/\psi$ events. By generating a $1.0\times10^{12}$ MC sample, after event selection and detector optimization, the statistical accuracy of $CP$ violation is $10^{-4}$. The STCF has great potential in improving luminosity and realizing beam polarization. It is expected that more than 1~ab$^{-1}$ experimental data and 3.4$\times10^{13}$ $J/\psi$ events will be obtained per year, with the substantial increase in statistics, larger data samples will be generated on STCF, and in the future, it will hopefully reach a level of accuracy and theoretical prediction compatibility.

\section{\boldmath Summary and prospect}

With the fast simulation software package, the MC samples of $J/\psi\to\Lambda\bar{\Lambda}\to p\pi^{-}\bar{n}\pi^{0}$ process were generated. After the optimization of detector performance, the events selection efficiency of the signal process is increased by $23.22\%$ compared to the unoptimized and the main background process is reduced by $5.5\%$. Furthermore, the $1.0\times10^{12}$ $J/\psi$ MC was used to pre-studied the sensitivity of $CP$ violation of the $J/\psi\to\Lambda\bar{\Lambda}$ process at the future STCF. The statistical accuracy of $CP$ violation of $\Lambda$ hyperon is $10^{-4}$, which is close to the prediction of SM of $CP$ violation in $\Lambda$ hyperon decay~\cite{predict Lam}. 

\section*{\boldmath Acknowledgments}
The authors would like to thank the USTC Supercomputing Center and the Hefei Comprehensive National Science Center for their strong support. This work is supported in part by National Natural Science Foundation of China~(NSFC) under Contracts Nos. 11872030, 11905092, 11972177, 12122509, 11625523, 12105132, 11705078. The international partnership program of the Chinese Academy of Sciences under Grant No. 211134KYSB20200057 and by USTC Research Funds of the Double First-Class Initiative and the Fundamental Research Funds for the Central Universities. The Doctoral Scientific Research Foundation of Liaoning Province No. 2019-BS-113, the Foundation of Liaoning Educational Committee No. LQN201902, the Natural Science Foundation of Liaoning Provincial Department of Education No. LCJ202003. China Postdoctoral Science Foundation under Contracts Nos. 2021M693181. The PhD Start-up Fund of Natural Science Foundation of Liaoning Province of China under Contracts No. 2019-BS-113. Scientific research Foundation of Liaoning Provincial Department of Education under Contracts No. LQN201902. Foundation of Innovation team 2020, Liaoning Province. Opening Foundation of Songshan Lake Materials Laboratory, Grants No.2021SLABFK04. 


\end{document}